\documentclass[conference]{IEEEtran}
\IEEEoverridecommandlockouts

\usepackage{graphicx}
\usepackage{comment}
\usepackage{amsmath}
\usepackage{amssymb}
\usepackage{array}
\usepackage{enumitem}
\usepackage[colorlinks,urlcolor=blue,linkcolor=blue,citecolor=blue]{hyperref}

\usepackage{hyperref}
\hypersetup{ 
 colorlinks=true, 
 linkcolor=blue, 
 filecolor=blue, 
 citecolor = blue, 
 urlcolor=blue, 
} 

\begin{document}

\title{Onion Routing Key Distribution for QKDN}

\author{\IEEEauthorblockN{Pedro Otero-García\IEEEauthorrefmark{1}, Javier Blanco-Romero\IEEEauthorrefmark{2}, Ana Fernández-Vilas\IEEEauthorrefmark{1}, Daniel Sobral-Blanco\IEEEauthorrefmark{2} \\ Manuel Fernández-Veiga\IEEEauthorrefmark{1},  Florina Almenares-Mendoza\IEEEauthorrefmark{2}}
\thanks{This work was supported under the grant TED–2021–130369B–C31 funded by MICIU/AEI/
10.13039/501100011033 and by the “European Union NextGenerationEU/PRTR” and the grant  PID2020–113795RB–C32 funded by MICIU/AEI/10.13039/501100011033.}
\IEEEauthorblockA{\IEEEauthorrefmark{1}atlanTTic, Universidade de Vigo \\
\{pedro.otero,avilas,mveiga\}@det.uvigo.es}
\IEEEauthorblockA{\IEEEauthorrefmark{2}\textit{Department of Telematic Engineering},
\textit{Universidad Carlos III de Madrid}, Leganés, Madrid, Spain \\
\{frblanco,dsobral\}@pa.uc3m.es, florina@it.uc3m.es}}

\maketitle

\begin{abstract}
The advance of quantum computing poses a significant threat to classical cryptography, compromising the security of current encryption schemes such as RSA and ECC. In response to this challenge, two main approaches have emerged: quantum cryptography and post-quantum cryptography (PQC). However, both have implementation and security limitations. In this paper, we propose a secure key distribution protocol for Quantum Key Distribution Networks (QKDN), which incorporates encapsulation techniques in the key-relay model for QKDN inspired by onion routing and combined with PQC to guarantee confidentiality, integrity, authenticity and anonymity in communication. The proposed protocol optimizes security by using post-quantum public key encryption to protect the shared secrets from intermediate nodes in the QKDN, thereby reducing the risk of attacks by malicious intermediaries. Finally, relevant use cases are presented, such as critical infrastructure networks, interconnection of data centers and digital money, demonstrating the applicability of the proposal in critical high-security environments.
\end{abstract}

\section{Introduction \label{sec:intro}}


The development of quantum computing is progressing at a steady pace, and it is becoming increasingly evident that quantum algorithms, such as Shor's~\cite{Shor1999} and Grover's~\cite{Grover1996}, will be efficiently implementable in the near future. This advancement will considerably affect the realm of computer security, as current classical cryptography, which is widely used in modern IT systems, will become vulnerable.

In particular, public-key algorithms, such as RSA, Diffie-Hellman and elliptic curve-based algorithms (ECC), will lose effectiveness because they are based on mathematical problems that a sufficiently powerful quantum computer could solve in a feasible amount of time. For example, Shor's algorithm can factor large numbers in polynomial time, completely breaking the security of RSA. Similarly, the discrete logarithm problem, the basis of ECC and Diffie-Hellman, also becomes ineffective in front of quantum attacks.

In the case of symmetric encryption algorithms such as AES and cryptographic hash functions such as SHA-2 and SHA-3, the quantum threat is also relevant, although less severe. Grover's algorithm can reduce the time required for an exhaustive search (brute force) attack from \(2^n\) to \(2^{n/2}\), which implies that, to maintain the same level of current security, the size of the keys would have to be doubled. In other words, to achieve the current level of security provided by AES-128, AES-256 would need to be employed.

\subsection{Response to the quantum threat \label{subsec:resp-th}}

Given this quantum scenario, there has been a growing interest in the scientific community to develop cryptographic algorithms that are resistant to quantum attacks. There are two main approaches: quantum cryptography and post-quantum cryptography.

Quantum cryptography relies on the properties of quantum mechanics to design secure algorithms that are secure against quantum attacks. A prominent example is quantum key distribution (QKD), which allows secure key exchange thanks to quantum properties such as superposition,  with the BB84~\cite{Bennett1984} algorithm being the most pioneer in this group, and quantum entanglement, where E91~\cite{Ekert1991} is highlighted.

Nevertheless, the implementation of QKD solutions depends on the development of quantum hardware, which represents a significant challenge. Although feasible, the current technology is expensive and requires direct fiber optic connections without repeaters, limiting its applicability to specific scenarios. Furthermore, QKD does not replace classical cryptography in its entirety, but rather focuses on improving the security of key exchange without providing solutions for asymmetric cryptography, to compute shared keys via KEMs (Key Encapsulation Mechanism), or cryptographic signatures.

Post-quantum cryptography adopts a more pragmatic approach to the quantum threat than its quantum counterpart. Rather than relying on specialized hardware, it seeks to develop algorithms based on mathematical problems that remain difficult to solve even with a quantum computer. Unlike quantum cryptography, these algorithms can be implemented on classical IT hardware, facilitating their integration into existing systems without the need for complex infrastructures or expensive devices.

The National Institute of Standards and Technology (NIST) is currently spearheading a process of standardization of post-quantum algorithms~\cite{Alagic2022}, with the objective of establishing the security standards that will be in place in the quantum era. In 2022, following the third round of its PQC (Post-Quantum Cryptography) initiative, the algorithms that have been selected for future standardization were announced. These include CRYSTALS-Kyber~\cite{avanzi2020crystals} (Kyber), a public-key encryption scheme, and CRYSTALS-Dilithium~\cite{lyubashevsky2020crystals} (Dilithium), a digital signature algorithm. Both are designed to ensure security even in the face of quantum attacks.

\subsection{Advantages and disadvantages of QKD \& PQC \label{subsec:qkd-pqc}}

At first glance, PQC has major advantages over quantum cryptography, since its real-world implementation is more immediate, versatile and economical. However, despite the hardware limitations of quantum computing, QKD remains the only option that has been proven to be completely secure against quantum attacks. Post-quantum cryptography, moreover, is susceptible to be vulnerable by both quantum and classical attacks that could still be discovered~\cite{zeng2024practical}. This may seem a bold statement, but it is not the first time this has occurred. Previously popular algorithms, such as classic MD5 hash function or 3DES and RC2 encryption schemes~\cite{OpenStackCrypto}, are nowadays considered insecure due to insufficient key sizes or the emergence of efficient attacks against them. In addition, side-channel attacks~\footnote{Side-channel attacks exploit vulnerabilities in the way algorithms or protocols are implemented.} are more likely against PQC algorithms than QKD algorithms, since PQC implementations are still under development.

Even though QKD is the most secure alternative, its limitation on key distribution between two unconnected nodes of the same network is undeniable. Currently, existing QKD protocols require two nodes to be directly connected by a quantum channel to compute a shared key. One of the main challenges for QKD networks (QKDNs), i.e., networks consisting of several interconnected QKD nodes, is the creation of a shared key between any pair of nodes within the QKDN. The most obvious solution would be for all nodes to share a quantum channel, but this option is not scalable. In fact, two nodes separated by a significant distance will not be able to generate a shared secret due to the distance limitations of quantum communication and the absence of quantum repeaters. There are currently two models that provide a solution to this problem, key-relay and trused-node. Nevertheless, both models do not respect the confidentiality end-to-end of the key by the intermediate nodes, as discussed in section~\ref{subsec:qkdn}.

\subsection{Objectives and structure of the article}

This article proposes a secure key distribution protocol in which any two nodes in a QKDN can establish a shared secret using the key-relay based key distribution model, complemented with the encapsulation philosophy of onion routing and post-quantum techniques. This approach attempts to guarantee the CIA triad (Confidentiality, Integrity and Authenticity), while introducing anonymity in the communication between nodes thanks to the onion encapsulation. Furthermore, classic key distribution problems in QKDN, such as scalability or anonymity, will be mitigated by combining QKD and PQC algorithms.

The rest of the document is organized as follows: Section~\ref{sec:tech-bg} presents 
the technical background used for the development of the project, introducing the 
current key distribution methods in QKDN and the basics of onion routing. Section~\ref{sec:qsn} describes the architecture and security model of the proposed network, justifying its robustness and explaining the proposed protocol. Section~\ref{sec:uoc}
details the use cases where the proposed model would be useful. Finally, Section~\ref{sec:cons} presents the conclusions of the article.

\section{Technical background \label{sec:tech-bg}}

\subsection{Key distribution methods in QKDN \label{subsec:qkdn}}

Currently, there are two common solutions for key distribution in QKDN: the key-relay~\cite{elliott2002building} implementation, also known as hop-by-hop, and the Trusted Node (TN)~\cite{ITU-T_Y3803} implementation, also known as Central KMS.

\begin{figure}[t]
    \centering
    \includegraphics[width=\columnwidth]{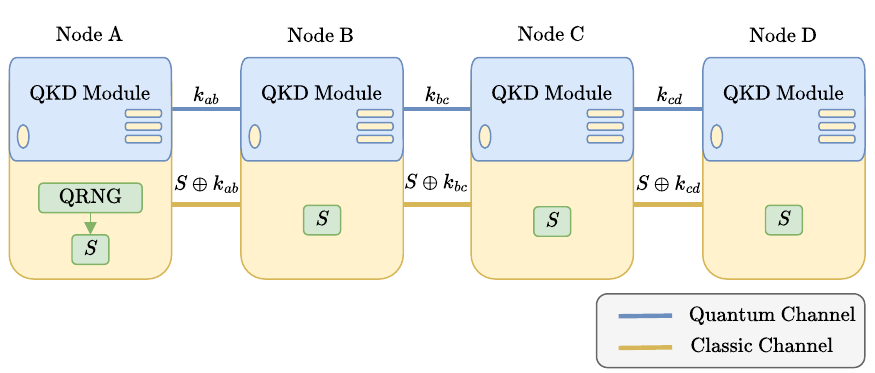}
    
    \caption{This figure shows an example of key relay operation. Node $A$ wants to share a shared secret $S$ with $D$, and their connection path pass through nodes $B$ and $C$. To do so, $A$ encrypts $S$ with the QKD key \(k_{ab}\) that it shares with $B$, and sends it to node $B$; node $B$ decrypts it, encrypts it with the QKD key \(k_{bc}\) that it shares with node $C$, and sends it to $C$. Finally, $C$ repeats the process and D decrypts the message that arrives from $C$ with the QKD key \(k_{cd}\) obtaining the shared secret $S$.
    \label{fig:key-relay}}
\end{figure}

In the key-relay approach, Figure~\ref{fig:key-relay}, two QKD nodes ($A$ and $D$) that wish to share a common secret ($S$) first determine a path of intermediate nodes between them. The initiating node then generates $S$ using a quantum random number generator (QRNG), encrypts it with the quantum key shared with its neighbor and transmits it over a classical channel. Each intermediate node decrypts the incoming ciphertext and repeats this process until the destination node recovers $S$. The main drawback of this method is that all intermediate nodes have access to $S$, which compromises the confidentiality of the shared secret. An example of this model is shown in Figure~\ref{fig:key-relay}.

\begin{figure}[t]
    \centering
    \includegraphics[width=\columnwidth]{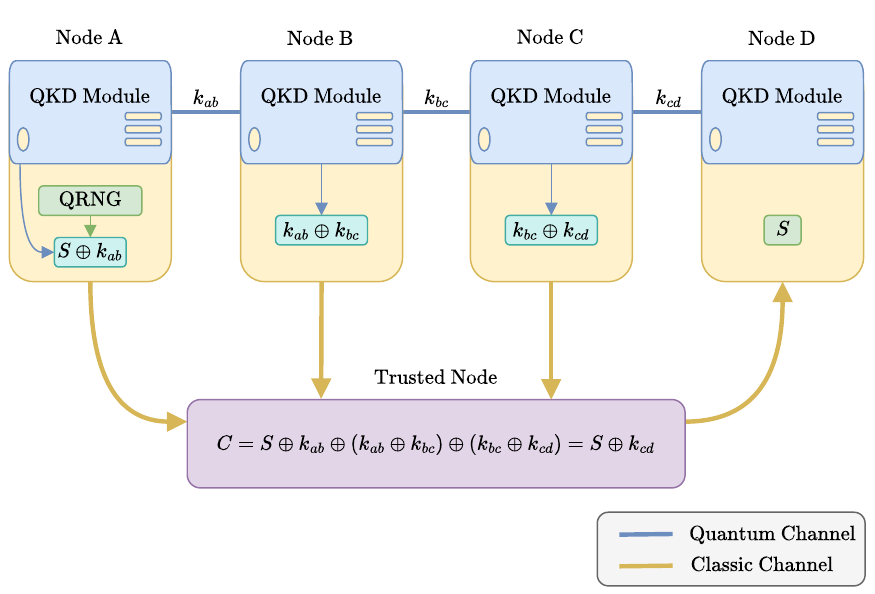}
    
    \caption{This diagram shows an example of TN operation. Node $A$ wants to share $S$ with node $D$ connected through the path that nodes $B$ and $C$ form. To do so, $A$ encrypts $S$ with the QKD key \(k_{ab}\) that it shares with its adjacent node $B$ and sends it via the classical channel to the TN. $B$ encrypts the key \(k_{ab}\) with the key \(k_{bc}\) that it shares with its other neighbor node $C$ and also sends it to the TN. The same procedure is done by $C$ with the key \(k_{bc}\). Since the encryption mechanism is an OTP (XOR), TN obtains $C = S \oplus k_{cd}$ which it sends to the final node D. This one is able to obtain S with the key $k_{cd}$ that it shares with its neighbor node $C$.
    \label{fig:tn}}
\end{figure}
\raggedbottom

In the Trusted Node (TN) model, Figure~\ref{fig:tn}, in addition to the nodes forming the QKDN, there is a TN that is trusted and reachable by all. Any node in the path between a pair ($A$ and $D$) that wants to share a secret $S$ provides cryptographic material to the TN through a classical channel, including the secret-generating node. This allows the TN to generate a ciphertext ($C$) from all the received material, which is then sent to the destination node. The latter can retrieve the secret $S$ from $C$ using a QKD key previously established with the last intermediate node of the path created. Figure~\ref{fig:tn} illustrates an example of the functionality employed by the TN model.

It is relevant to note that the TN does not necessarily have to be an external element to the quantum network, but can also coincide with the destination node ($D$) in the key sharing process. Nonetheless, as already occurred in the key-relay model, the TN-based model does not guarantee the confidentiality of $S$ between $A$ and $D$. This is because if one of the intermediate nodes ($B$ and $C$) capture the messages that reach the TN through the classical channel, they will have the necessary cryptographic material to obtain $S$ taking advantage of the OTP (One-Time Pad) encryption method to encrypt $S$ and the other encryption keys.

It is also important to acknowledge the potential for spoofing attacks on the classical channel in both implementations. Such attacks may be initiated by an intermediate node, the end node, or even the TN in this model. While it is unlikely that an attacker would obtain the quantum key necessary to decrypt the messages, since it would have to take control of a QKD node in order to take the quantum keys, the possibility of network disruption remains a concern.

\subsection{Brief definition of onion routing \label{sec:onion}}

\begin{figure}[t]
    \centerline{\includegraphics[width=\columnwidth]{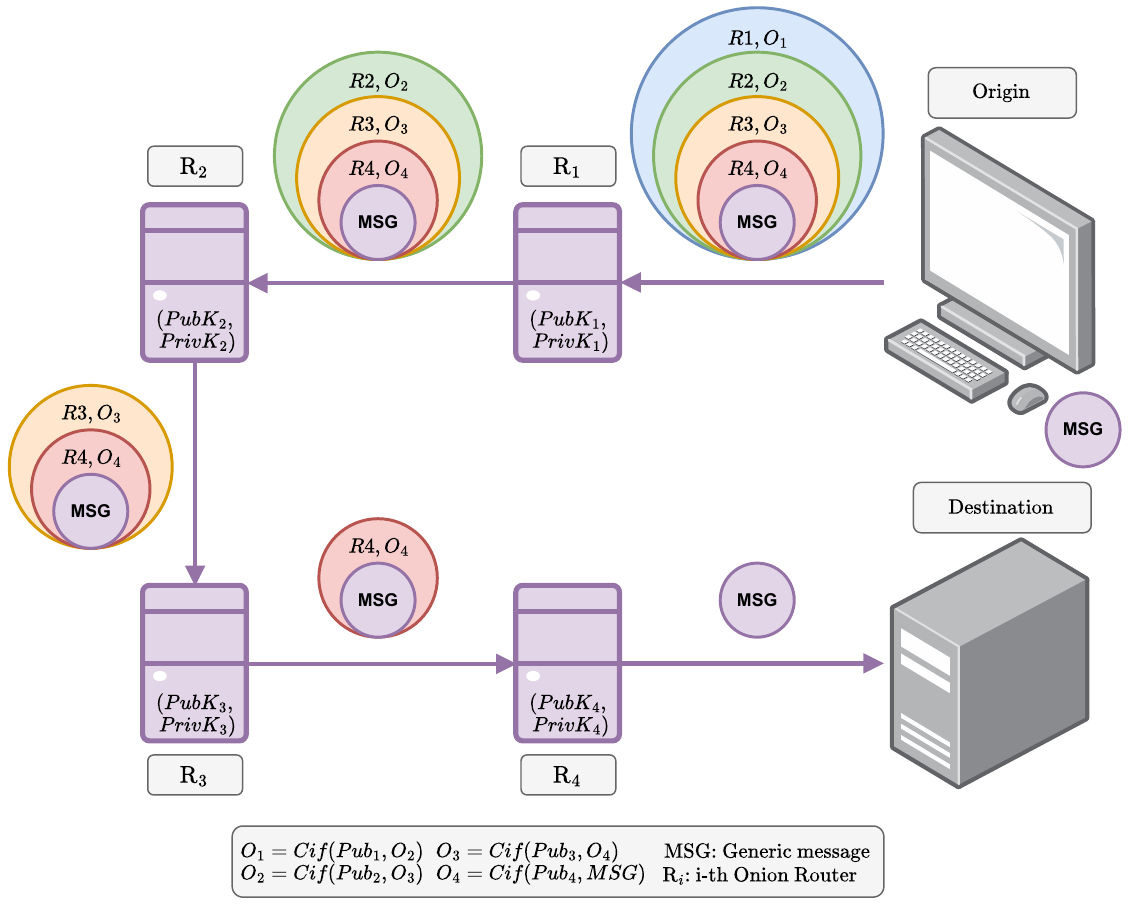}}
    \caption{Example of the encapsulation in an onion route circuit. Here the origin device makes an encapsulation encryption with all the public keys of the onion routers (OR) in the circuit and sends the ciphertext, also known as onion, through the circuit. Each OR decrypt the last encryption layer with its private key and sends the result to the next OR. The last OR router gets the original message and sends it to its destination.
    \label{fig:onion}}
\end{figure}

Onion Routing~\cite{goldschlag1999onion} (OR) is a method of anonymous communication designed to protect the identity and privacy of users on the Internet. Its operation is based on layered encryption, where a message is encrypted multiple times, forming the "onion encryption", before being sent through a series of intermediate nodes, called onion routers that forms a path called circuit. Figure~\ref{fig:onion} shows an example of onion encryption. Each node in the network only knows the previous and next node in the circuit, which prevents any single entity from having access to the complete sender and receiver information. The process begins with the selection of a path of intermediate nodes, after which the message is encrypted in successive layers, starting with the public key of the last node and progressing to the first. As the message traverses the network, each node decrypts its corresponding layer and forwards the remaining content to the next node, until finally the message reaches the exit node, where the last layer of encryption is removed and delivered to the final destination. This approach offers a high level of anonymity, since no intermediate node can know both the source and destination of the message.

OR has a large number of variants and extensions that improve security by adapting the protocol to specific scenarios. Among the most interesting extensions when sending messages are those that implement integrity~\cite{kuhn2020breaking,kuhn2021onion} between the initiator of the circuit and the onion routers. Thanks to integrity, the destination node can guarantee that the incoming message has not been tampered with and can prevent attacks such as message forwarding that can cause overloads on onion routers or even disable them.

Nonetheless, the implementation of OR introduces certain limitations, including increased communication latency and the potential vulnerability of outgoing nodes to surveillance or attacks. OR is extensively utilized in networks such as Tor (The Onion Router), a system that facilitates anonymous browsing, censorship circumvention, and online privacy protection. This functionality is particularly advantageous for professionals, including journalists and activists, as well as users with particular confidentiality needs in their communications.

\raggedbottom

\section{Proposal of QKDN protocol} \label{sec:qsn}


From the information presented in the previous sections, it is evident that both PQC and QKD have their own advantages and disadvantages. While PQC is notable for its versatility and applicability in conventional environments, QKD provides a superior level of security. In light of this dichotomy, the scientific community has initiated the development of hybrid solutions that integrate both technologies, thereby optimizing both security and implementation feasibility. A notable proposal entails the incorporation of PQC signature algorithms, such as Dilithium, to authenticate QKD nodes in key-relay and TN models\cite{james2023key}. This improvement enables the verification of messages transmitted through classical channels, which facilitates the detection of Man-in-the-Middle (MitM) attacks and prevents spoofing attempts.

Nonetheless, uncertainty remains with regard to the protection of the shared secret from the intermediate nodes in both QKDN models analyzed (key-relay and Trusted Node). Just as PQC signatures for message authentication, PQC algorithms, in this case public key KEM such as Kyber, can also be employed to ensure the confidentiality of messages transmitted by QKD nodes over a classical channel. If PQC encryption were applied correctly, it would prevent intermediate QKD nodes from being able to obtain the shared secret between the end nodes during their transmission.

In the key-relay scheme, the implementation of PQC public key encryption is relatively simple. The initiating node encrypts the shared secret $S$ with the destination node's symmetric key computed previously with a PQC-KEM, keeping the rest of the procedure unchanged. Thus, when an intermediate node decrypts the ciphertext with the QKD key shared with its previous node, it does not get $S$ in plaintext, but $S$ encrypted with the symmetric key shared by destination and initiator. Finally, the destination node decrypts the received ciphertext with its QKD key and, subsequently, with its PQC-KEM key to recover $S$. 

On the other hand, in the TN model, the incorporation of PQC-KEM is also straightforward. Rather than transmitting its cryptographic material in clear form, each node employs encryption with a symmetric PQC-KEM key shared with the TN. Subsequently, the TN, upon receiving the encrypted messages, employs decryption with the corresponding PQC-KEM key, calculates the ciphertext $C$, encrypts it with symmetric key shared with the destination node, and forwards it. The destination node, upon receiving $C$, employs the PQC-KEM key that shares with TB to decrypt it and thereby obtain the shared secret.

In terms of performance, the incorporation of PQC-KEM results in increased latency in the TN architecture relative to key-relay. In key-relay, a single encryption and decryption process with PQC is sufficient, whereas in TN, this procedure must be repeated for each node. If the TN corresponds to the end node, one iteration can be eliminated; however, in scenarios involving multiple intermediate nodes, this optimization is negligible.
Regardless of the QKD network model selected, hybridization with PQC-KEM encryption provides the robustness of QKD key-based encryption along with the flexibility of PQC. Therefore, in the event of an intermediate node in the QKD network being compromised, the shared secret will remain protected and will not be exposed. Conversely, in the event of a vulnerability in the PQC algorithm employed, the shared secret remains inaccessible to external attackers and is exclusively discernible by the QKD nodes within the network, thus preserving the security of the original system without hybridization.

\subsection{Proposed protocol \label{sec:alg}}

\begin{figure*}[t]
    \centerline{\includegraphics[width=\textwidth]{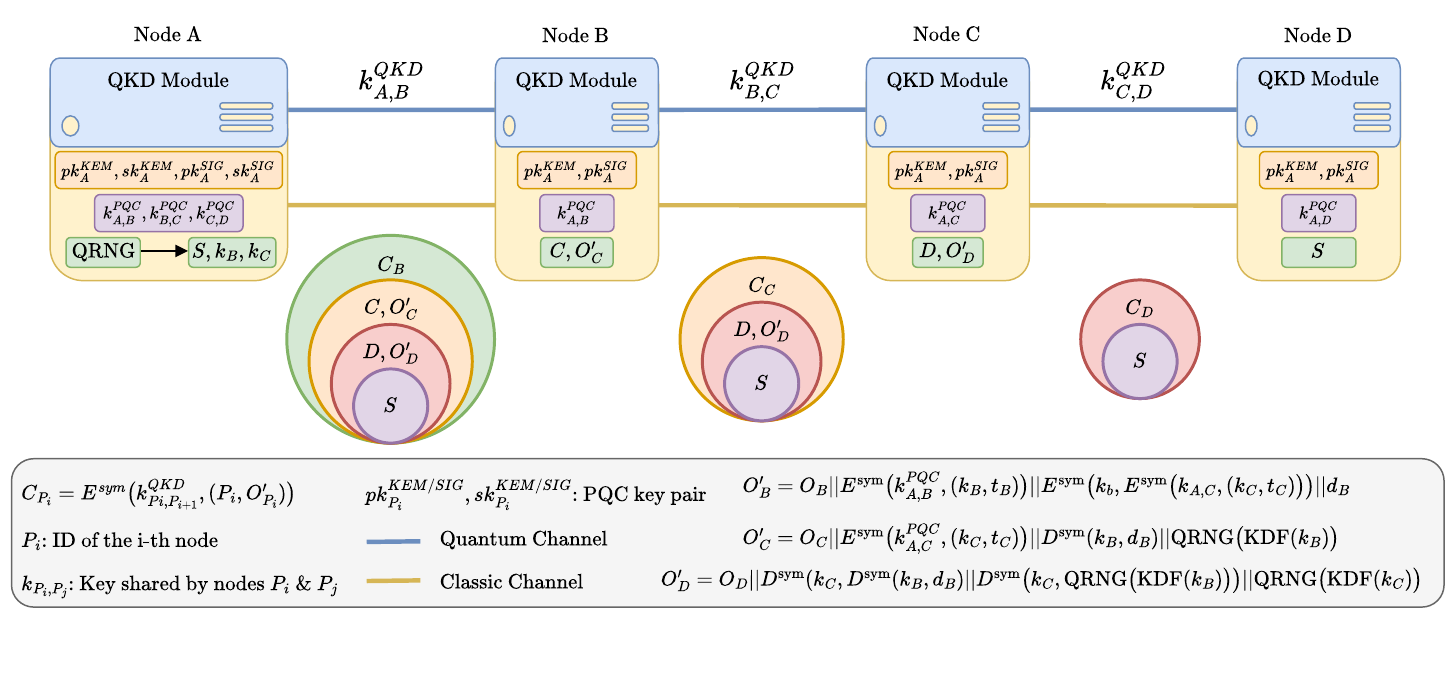}}
    \caption{Operation diagram for the proposed protocol to ensure communication of the secret between two nodes within a QKDN while maintaining confidentiality, integrity, authentication and anonymity. The signature keys are used in the process of creating and verifying $t_{P_i}$.
    \label{fig:qsn-key-dist}}
\end{figure*}
\raggedbottom

The hybrid QKD model with PQC signature and PQC-KEM almost ensures compliance with the principles of the CIA triad (Confidentiality, Integrity and Authenticity), only integrity needs to be guaranteed. It is true that integrity may be implicit in the model since commercial QKD nodes, such as Cerberis XGR QKD by ID Quantique~\cite{nodes-idq}, already guarantee integrity between pairs of nodes in the key sharing process, but not in external messages to the QKD protocol. 

Furthermore, anonymity should also be taken into account since it has become increasingly important in the field of network security. Achieving integrity and anonymity in a conventional network is not a trivial task, although there are widely adopted solutions for this purpose, the most prominent being the TOR network based on an extended version of OR.


Based on the multi-hop behavior of the key-relay model and its similarities with OR, we have developed the following proposal for a Secure QKDN Protocol that not only complies with the CIA triad, but also guarantees anonymity among nodes. The design of the protocol is grounded on the following assumptions:

\begin{enumerate}
    \item The QKDN will be composed of an undetermined number of QKD nodes, capable of negotiating quantum keys with one or more nodes within the network.
    
    \item Each QKD node will be able to exchange message through a classical channel with the rest of the nodes within the network.
    
    \item The neighboring QKD nodes will generate QKD shared keys in advance in order to optimize communication and minimize delays.
    
    \item For any pair of randomly selected QKD nodes within the network, there must be at least one path (circuit) of nodes connecting them.
    
    \item Each node will have a key pair consisting of a public key and a private key to encrypt/decrypt messages base in a PQC-KEM algorithm, e.g. Kyber.  
    
    \item Each node will have the ability to sign messages and verify the signs of other nodes using a post-quantum digital signature algorithm, e.g. Dilithium.
    
    \item The public keys, for public encryption or signature schemes, of any node can be obtained by other nodes through the classic communication channel.
\end{enumerate}

The following sections details the steps to create and processing the ciphertexts, onions, of the QKDN protocol proposal, based on~\cite{kuhn2021onion}. For this scenario, it is assumed that all the above-mentioned assumptions are met, that the initiator node has previously computed a PQC-KEM key with the other nodes in the network and all the public keys were distributed among the nodes. For a better understanding,  Figure~\ref{fig:qsn-key-dist} shows the messages exchanged in the protocol and Table~\ref{tab:nomenclature} shows the used nomenclature where $D^{\text{sym}}(k, E^{\text{sym}}(k, m)) = m$, and $\operatorname{Ver}\bigl(pk_i,\bigl(t, \operatorname{Sig}(sk_i,t)\bigr)\bigr)= 1$. Additionally, $(pk_i,sk_i)$ for encryption and signing have the same nomenclature but are different key pairs.

\begin{table}[t]
    \centering
    \renewcommand{\arraystretch}{1.3}
    \caption{Nomenclature used in the system \label{tab:nomenclature}}
    \begin{tabular}{cp{0.65\columnwidth}}
        \hline
        \textbf{Nomenclature} & \textbf{Description} \\
        \hline
        $n$ & Number of intermediate QKD nodes in the circuit\\
        \hline
        $N$ & $n+1$, including destination node \\
        \hline
        $P_i$ & Identity of $\text{i-th}$ \\
        \hline
        $\mathsf{pk}_i$ & Post-quantum public key of the $i$-th QKD node \\
        \hline
        $\mathsf{sk}_i$ & Post-quantum private key of the $i$-th QKD node \\
        \hline
        $k_i$ & Symmetric key shared by the initiator and node $i$\\
        \hline
        $k_{P_i,P_{i+1}}$ & QKD key shared by the nodes $P_i$ and ${P_{i+1}}$ \\
        \hline
        $t_i$ & Node $i$'s MAC tag \\
        \hline
        $O_i$ & Original onion for $i$ \\
        \hline
        $O^\prime_i$ & Extended onion for $i$ \\
        \hline
        $\mathsf{rdm}_i$ & Padding for node $i$ \\
        \hline
        $a || b$ & Concatenation of bit strings $a$ and $b$ \\
        \hline
        $(P_1,\dots,P_{n+1})$ & Circuit of onion routers ($P_0$ is the initiator node)\\
        \hline
        $E^{\text{sym}}(k, \cdot)$ & Symmetric key $k$ encryption with AES-256-CBC \\
        \hline
        $D^{\text{sym}}(k, \cdot)$ & Symmetric key $k$ decryption with AES-256-CBC \\
        \hline
        $\mathsf{Sig}(\mathsf{sk}_i,\cdot)$ & Sign with Dilithium \\
        \hline
        $\mathsf{Ver}\bigl(\mathsf{pk}_i,(\cdot,\cdot)\bigr)$ & Verify sign with Dilithium \\
        \hline
        $H(\cdot)$ & Compute hash SHA-2 \\
        \hline
    \end{tabular}
    
\end{table}
\raggedbottom

\emph{1) Creation of new ciphertext}
\label{subsubsec:creation-on}
\begin{enumerate}[label=\roman*.] 
    \item Generate the one-time keys $k_1, \dots, k_n$, one for each node in the circuit, and $S$ with a QRNG.
    
    \item Compute original onions: For the first encryption in the encapsulation 
    $O_{n+1} = (P_{n+1} || E^{\text{sym}}(\mathsf{k}^{PQC}_{P_0,P_N},S) )$ and $O_i = \bigl(P_i || E^{\text{sym}}(\mathsf{k}^{PQC}_{P_0,P_i},O_{i+1})\bigr)$ for $i = n, \dots, 1$.
      
    \item Compute encrypted dummy paddings $(\mathsf{rdm}_1, \dots, \mathsf{rdm}_n)$:
        \begin{enumerate}
            \item Select $N$ padding blocks $(B_1, \dots, B_N)$ at random and with the same length.
            \item Simulate the process for each intermediate node. For $i =
              1, \dots, n$
                \begin{enumerate}
                    \item Store the block that are not going to be replaced with tags. \(\mathsf{rdm}_i = (B_{n - (i-1)}, \dots, B_N)\).
        
                    \item Compute each block as $i$-th router would do, i.e. replacing a block with a tag: $r_i \leftarrow \mathsf{QRNG}\bigl( \mathsf{KDF}(P_i, k_i)\bigr)$ and
                        $(B_1,\dots,B_N) \leftarrow \bigl( D^{\text{sym}}(k_i, B_2), \dots, D^{\text{sym}}(k_i, B_N), r_i \bigr)$.
                \end{enumerate}
        \end{enumerate}
      
    \item Compute integrity tags $(t_1, \dots, t_n)$: for $i = n, \dots,1$:
        \begin{enumerate}
                \item Store the tags until here: $t_i \leftarrow (B_1,
                \dots, B_{n-i})$
                
                \item Encrypt all blocks 
                \begin{equation*}
                    (B^\prime_1, \dots, B^\prime_{N-1}) \leftarrow \bigl(
                  E^{\text{sym}}(k_i, B_1), \dots, E^{\text{sym}}(k_i, B_{N-1}) \bigr)
                  \end{equation*}
                
                \item Generate the tag $\tau_i$ and embed it to the key: 
                    \begin{enumerate}
                        \item $\tau_i \longleftarrow \mathsf{Sig}(sk_i, (O_{i+1} || B^\prime_1 || \cdots || B^\prime_{N-1}))$
                        
                        \item $B_{\text{new}} \leftarrow E^{\text{sym}}(\mathsf{k}_{i},\mathsf{embed}(k_i, \tau_i))$
                    \end{enumerate}
                
                \item Replace the blocks.
                \begin{equation*}
                    (B_1, \dots, B_N) \leftarrow (B_{\text{new}}, B^\prime_1,
                  \dots, B^\prime_{N - 1}).
                  \end{equation*}
        \end{enumerate}
      
    \item Combine the extensions created in the previous steps with the original onions
    \begin{equation*}
    (O^\prime_1, \dots, O_{n+1}^\prime) \leftarrow (O_1 || t_1 ||
        \mathsf{rdm}_1, \dots, O_{n+1} || t_{n+1} || \mathsf{rdm}_{n+1}).
        \end{equation*} 

    \item Then, it encrypts \(O'_1 = (O_1||t_1||\mathsf{rdm}_1)\) with the QKD key \(k^{QKD}_{P_0,P_1}\)\footnote{The algorithm for this symmetric encryption will be AES achieving semantic security as OTP is not feasible due to the length of the delivered message.} and sends the ciphertext to $P_1$\footnote{It is not necessary to sign the $O'_i$ ciphertext since they are encrypted with a symmetric QKD key and when these are negotiated they already guarantee the integrity between a pair of nodes.}.
\end{enumerate}

\emph{2) Processing incoming ciphertext}
\label{subsubsec:processing-on}
\begin{enumerate}[label=\roman*.] 
    \item Decrypt the incoming ciphertext with $k^{QKD}_{P_{i-1},P_{i}}$ getting the onion \(O'_i\).
    
    \item Split \(O'_i\) to obtain the original onion and its extension \((O_i, \mathsf{ext}_i)\).

    \item Process classic onion obtaining: 
        \begin{enumerate}
            \item \((P_{N}, S) = (P_{N} || D^{\text{sym}}(\mathsf{k}^{PQC}_{1,P_N},S) )\)

            \item \((P_{i+1}, O_{i+1})= (P_i || D^{\text{sym}}(\mathsf{k}^{PQC}_{1,P_{i+1}},O_{i+1}))\) 
        \end{enumerate}

    \item Split the extension in the blocks \((B_1, \dots, B_N)\)

    \item If receptor, return $(\emptyset, O_{i+1})$

    \item Extract symmetric key and tag. \((k_i, t_i) \longleftarrow \mathsf{Extract}\bigl(
        D^{\text{sym}}(\mathsf{k}_{1,i},B_1) \bigr)\)

    \item Check tag $t_i$. If it is invalid, discard the onion. \(\mathsf{Ver}_{pk_i}\bigl(t_i, (O_{i+1}||B_2|| \cdots || B_N)
        \bigr) == 1\)

    \item Generate new padding:
        \begin{enumerate}
            \item \(r_i \longleftarrow \mathsf{QRNG}\bigl( \mathsf{KDF}(P_i, k_i)\bigr)\)
            
            \item \((B_1,\dots,B_N)\longleftarrow \bigl( D^{\text{sym}}(k_i, B_2), \dots, D^{\text{sym}}(k_i, B_N), r_i \bigr)\) 

            \item \(O^\prime_{i+1} \longleftarrow O_{i+1} || (B_1,\dots,B_N)\)
        \end{enumerate}

    \item Then, encrypt $(P_{i+1}, O^\prime_{i+1})$ with QKD key $k^{QKD}_{P_i,P_{i+1}}$ and send it to \(P_{i+1}\).
\end{enumerate}

\section{Use Cases \label{sec:uoc}}

This section details some use cases in which the fulfillment of the CIA triad with anonymity is fundamental and, as a consequence, it would be realistic to implement a QKDN network. These scenarios are characterized by the presence of extensive private networks and the economic capacity to adopt quantum nodes in the pre-quantum era. The following are some of the most notable examples:

\begin{itemize}
    \item \textbf{Protecting critical infrastructure}: Power grids, government agencies, telecommunications infrastructure and terrestrial networks depend on secure link communications to operate. The use of the proposed QKDN protocol would prevent cyberattacks that could compromise essential services and affect the stability of key sectors.

    \item \textbf{Communication inter-data centers}: Secure data transfer between data centers is crucial for content distribution, backups, distributed training of AI/ML models and execution of digital twins. Quantum cryptography would ensure the confidentiality and resilience of these processes against advanced attacks.

    \item \textbf{Digital Currencies and Financial Transactions}: The digitization of money requires robust mechanisms against fraud and cyberattacks. The implementation of quantum technologies in banks and digital payment systems would ensure the authenticity and integrity of transactions, thus protecting trust in the financial system.
\end{itemize}

\section{Conclusions \label{sec:cons}}

The growing threat of quantum computing on current cryptographic systems makes it imperative to develop solutions that guarantee the security of communications in the post-quantum era. In this work, a hybrid protocol has been proposed that combines QKD with PQC and the OR concept to provide a secure and anonymous solution in quantum communication networks. This approach allows protecting the confidentiality of shared secrets even in the presence of compromised intermediate nodes, while ensuring authenticity and integrity in data transmission. The use of encapsulation techniques inspired by OR, together with the integration of post-quantum public-key encryption, allows QKDN to implement anonymity between the QKD nodes that wants to shared a secret. Although the implementation of this protocol may involve higher latency compared to other models, its performance is not far from classic anonymous networks such as TOR. The proposed protocol will be implemented in a real QKDN in order to conduct a security and performance analysis to evaluate its feasibility.

\end{document}